\documentclass[12pt]{iopart}
\usepackage{iopams}
\usepackage{graphicx}
\usepackage{color}
\usepackage[pdftex,colorlinks,bookmarks=false,citecolor=blue,linkcolor=red,urlcolor=blue]{hyperref}

\begin{document}
\def\bea{\begin{eqnarray}}
\def\eea{\end{eqnarray}}
\def\be{\begin{equation}}
\def\ee{\end{equation}}
\def\Q{\mathbf{Q}}
\def\et{{\it et al.}}
\title{Dynamical dimer correlations at bipartite and non-bipartite Rokhsar-Kivelson points 
}
\author{Andreas M.~L\"auchli}
\address{Institut Romand de Recherche Num\'erique en Physique des 
  Mat\'eriaux (IRRMA), CH-1015 Lausanne, Switzerland}
\ead{laeuchli@comp-phys.org}
\author{Sylvain Capponi}
\address{Laboratoire de Physique Th\'eorique, IRSAMC, CNRS, Universit\'e Paul Sabatier, F-31062 Toulouse, France}
\author{Fakher F.~Assaad}
\address{Institut f\"ur Theoretische Physik und Astrophysik, Universit\"at W\"urzburg, Am Hubland, D-97074 W\"urzburg, Germany}
\date{\today}
\begin{abstract}
We determine the dynamical dimer correlation functions of quantum dimer models
at the Rokhsar-Kivelson point on the bipartite square and cubic lattices
and the non-bipartite triangular lattice. Based on an algorithmic idea by Henley, we simulate a 
stochastic process of classical dimer configurations in continuous time and perform
a stochastic analytical continuation to obtain the dynamical correlations in momentum space
and the frequency domain. This approach allows us to observe directly the dispersion relations and the
evolution of the spectral intensity within the Brillouin zone beyond the single-mode 
approximation. On the square lattice, we confirm analytical predictions related to soft
modes close to the wavevectors $(\pi,\pi)$ and $(\pi,0)$ and further reveal the existence 
of shadow bands close to the wavevector $(0,0)$. 
On the cubic lattice the spectrum is also gapless but here only a single 
soft mode at $(\pi,\pi,\pi)$ is found, as predicted by the single mode approximation. 
The soft mode has a quadratic dispersion at very long wavelength, but crosses over to a
linear behavior very rapidly. We believe this to be the remnant of the linearly dispersing "photon" 
of the Coulomb phase.
Finally the triangular lattice is in a fully gapped liquid phase where the bottom of the 
dimer spectrum exhibits a rich structure. At the $M$ point the gap is minimal and the
spectral response is dominated by a sharp quasiparticle peak. On the other hand, at the $X$ 
point the spectral function is much broader. We sketch a possible explanation based on the crossing
of the coherent dimer excitations into the two-vison continuum.
\end{abstract}
\maketitle

\section{Introduction}
Quantum Dimer models (QDM) are important minimal models for the understanding
of topological order, spin liquid phases and deconfinement. 
The discovery of a topologically ordered resonating valence-bond (RVB) phase of the QDM on the triangular 
lattice~\cite{Moessner01} has led to an intense exploration of the properties of dimer models 
in two and three dimensions on various lattices~\cite{Moessner01a,Misguich02,Ivanov04,Huse03,Moessner03} 
and recently also the doping process has regained a lot of 
attention~\cite{Rokhsar88,Balents05,Poilblanc06}.

While many ground state properties of QDMs are fairly well understood by now, the 
excitation spectrum has not yet been explored to the same depth. Many basic features of
the excitation spectrum are only known  at the level of the Single Mode Approximation (SMA), which
gives exact results on the presence of gapless modes.
However the actual functional form of dimer and other spectral functions are not known in great
details. 

Recently, first steps towards a quantitative description of the excitation spectrum of the triangular 
lattice QDM have been undertaken, based on numerical measurements of excitation gaps of dimer 
and vison excitations at~\cite{Ivanov04} and away~\cite{Ralko06,Ralko07}  from the 
Rokhsar-Kivelson (RK) point. The aim of the present study is to go one step beyond by determining the full energy 
resolved dynamical dimer spectral function of QDMs at the RK point on three different Bravais lattices 
in two and three dimensions.

In two dimensions there are two different classes of phase diagrams of 
QDMs, those for bipartite lattices, such as the square and the honeycomb lattice
and those for non-bipartite lattices such as the triangular lattice. 
The bipartite lattices generically allow a mapping onto a height model~\cite{Blote82}
and its subsequent analysis~\cite{Henley97} turns out to yield critical dimer-dimer correlations
and gapless modes at the RK point. The non-bipartite lattices instead do not allow such a mapping, 
and the triangular lattice for example has been shown to possess an extended dimer-liquid ground state 
characterized by topological order and a finite gap to excitations~\cite{Moessner01,Moessner01a,Fendley02, Ioselevich2002}.

In order to highlight the excitation spectra in the different cases, we choose to analyze the dynamical dimer 
correlations on the square and the triangular lattice in two dimensions, as well as the three-dimensional simple cubic lattice.

The plan of our paper is the following. We describe the numerical algorithm, the relevant observables 
and the stochastic analytic continuation method used in \Sref{sec:algo_obs}.
Then we present our main numerical results obtained for the square lattice (\Sref{sec:square})
and the cubic lattice (\Sref{sec:cubic}), followed by the triangular lattice (\Sref{sec:triangular})
and we conclude in \Sref{sec:conclusion}.

\section{Hamiltonian, Observables and the Algorithm}
\label{sec:algo_obs}
In the following, we study the simplest versions of QDMs on three different Bravais lattices
(square, cubic, triangular). The Hamiltonian is defined as follows:
\begin{equation}
H_{QDM} = \sum_\mathrm{\Box} -t\ \Box_\mathrm{Flip}  + V\ \Box_\mathrm{Flippable}
\end{equation}
where the sum runs over all plaquettes of four sites formed by nearest-neighbor bonds only,
the $\Box_\mathrm{Flip}$ operator flips two parallel dimers
on a plaquette, while $\Box_\mathrm{Flippable}$ is a diagonal term which is equal to one if the plaquette 
is flippable, i.e. has two parallel dimers, and is zero otherwise. 
In the following, we focus on the RK point ($t$ = 1, $V$ = 1) on all lattices, where the equal amplitude
superposition of all dimer coverings is a ground state~\cite{Rokhsar88}.

In Ref.~\cite{Henley04}, Henley has proposed a continuous-time Monte Carlo algorithm based on local plaquette-flip updates 
for classical dimers, in which the Monte Carlo time maps to the imaginary time axis of the corresponding QDM 
at the RK point. Therefore by calculating Monte Carlo time displaced correlation functions,  
one equivalently measures the imaginary time correlation functions of the quantum dimer 
model at the RK point. This algorithm turned out to be a special case of the more general continuous time 
Diffusion Monte Carlo algorithm~\cite{Syljuasen05}, where the (quantum) Monte Carlo dynamics becomes classical if the 
ground state wave function is known exactly, as it is the case at the RK point.

This algorithm and its extensions have already been used in the past to extract excitation gaps based on a simple fit 
of the asymptotic long-time behaviour of the imaginary time correlation functions, both for dimer- and vison-like excitations 
on the triangular lattice~\cite{Ivanov04,Ralko06,Ralko07}.

In this work, we now address for the first time the full energy and momentum resolved spectral response of dimer excitations on the 
square, the cubic and the triangular lattice by performing a stochastic analytical continuation of the imaginary time data collected this way.
This procedure allows us to reveal many interesting and rich features of the spectral functions, which are beyond hydrodynamic treatments.

\subsection{Observables}

In the following, we focus on the momentum resolved dynamical dimer correlation functions. 
These correlation functions are not only fundamental observables of any quantum dimer model,
but are also potentially relevant for experiments on spin models which map onto quantum dimer 
models~\cite{Moessner03b,Cabra04,Bergman06}.

We start by defining the momentum dependent dimer density operator: 
\be
\hat{D}_x(\Q):=\frac{1}{\sqrt{N}} \sum_{i}\ \exp[-\rmi \Q \cdot \mathbf{r}_i]\ n_x(i),
\label{eqn:resonon_structure}
\ee
where $N$ denotes the number of sites, $\Q$ is a given momentum, $\mathbf{r}_i$ 
labels the site coordinates and $n_\alpha(i)$ counts the number of dimers emanating in the $\alpha$ direction from site $i$ . 
We limit ourselves in this paper to a given dimer direction, here the $x$ axis on all lattices.
The equal-time dimer structure factor then reads:
\be
D_x(\Q):=|\hat{D}_x(\Q)|\mathrm{GS}\rangle|^2,
\ee
$|\mathrm{GS}\rangle$ being the ground state of the Hamiltonian. 
Finally we define the dynamical structure factor as
\be
\mathcal{D}_x(\Q,\omega):=-\frac{1}{\pi}\Im \lim_{\eta \rightarrow 0^+} \langle\mathrm{GS}|\hat{D}_x(-\Q)
\, \frac{1}{\omega-H+\rmi\eta} \,
\hat{D}_x(\Q)|\mathrm{GS}\rangle,
\label{eqn:dynd}
\ee
inserting a resolution of the identity gives
\be
\mathcal{D}_x(\Q,\omega)=\sum_{s} |\langle s|\hat{D}_x(\Q)| \mathrm{GS}\rangle|^2 \times \delta(\omega-E_s),
\label{eqn:dynd2}
\ee
where $s$ labels all the eigenstates with energy $E_s$ ($E_\mathrm{GS}$ is set to zero). 
Obviously we have $D_x(\Q)=\int_0^\infty \mathrm{d}\omega\ \mathcal{D}_x(\Q,\omega)$.
A prominent role will also be played by the first energy moment of the normalized spectral function
\be
\omega^{D_x}_\mathrm{fm}(\Q):=\frac{\int \mathrm{d}\omega\ \omega \mathcal{D}_x(\Q,\omega)}{D_x(\Q)}
=\frac{\sum_{s}\ E_s\ |\langle s|\hat{D}_x(\Q)| \mathrm{GS}\rangle|^2}{D_x(\Q)}
\ee

\subsection{Imaginary time correlations and analytical continuation}

The quantity which is actually measured in the classical Monte Carlo simulations 
is the Laplace transform of \eref{eqn:dynd}, resp. \eref{eqn:dynd2}:
\bea
\mathcal{D}_x(\Q,\tau)&=&\int_{0^-}^{\infty} \rmd \omega \exp[-\omega \tau] 
\ \mathcal{D}_x(\Q,\omega)\\
&=&\sum_{s}  |\langle s| \hat{D}_x(\Q)|\mathrm{GS}\rangle|^2 \times \exp[- E_s \tau].
\eea
The difficult part is then to recover the spectral weights $|\langle
s|\hat{D}_x(\Q)| \mathrm{GS}\rangle|^2$ and the excitation energies
$E_s$, based on Monte Carlo data which inevitably
contain error bars. This is a well-known example of an
ill-conditioned problem and originates from the fact that time-displaced
observables are obtained in imaginary time. In the Monte Carlo
community, this problem has been addressed to some extent by enforcing a Maximum Entropy
criterion to get dynamical spectra~\cite{MaxEnt}. The presence of an entropy term in the minimization 
procedure allows one to obtain the most probable spectrum given the Monte Carlo 
data and has been shown to be reasonably accurate in many situations.
In this work, we use a sophisticated formulation called Stochastic Analytic Continuation (SAC) initiated 
by Sandvik~\cite{Sandvik98} and formalized in detail by Beach~\cite{Beach04a},  
where instead of obtaining a single spectrum, many ``acceptable'' spectra are averaged.
There is some freedom left in the averaging procedure, but we emphasize that our findings and conclusions
are robust with respect to this freedom. For instance, although sum-rules can be enforced in the Maximum Entropy procedure, we 
refrain to do this, and instead we systematically check that the first moment of the spectral distribution obtained by SAC
is identical (within the error bars)  to the first moment obtained directly by the Monte Carlo simulations by virtue of the relation
$\omega^{D_x}_\mathrm{fm}(\Q)= - \frac{\mathrm{d}}{\mathrm{d}\tau} \mathcal{D}_x(\Q,\tau)/D_x(\Q) $. Via this procedure we
obtain an independent assessment of the quality of the stochastic analytical continuation.

\subsection{Single mode approximation (SMA)}

On the analytical side, the single mode approximation (SMA) has proven to be a
useful tool for the analysis of some aspects of the excitation spectrum of QDMs. 
Based on its definition
\be
\omega^{D_x}_\mathrm{SMA}(\Q):=\frac{1}{2}
\frac{\langle \mathrm{GS} | [\hat{D}_x(-\Q),[H,\hat{D}_x(\Q)]] |\mathrm{GS} \rangle}
{D_x(\Q)},
\ee
and the fact that this quantity constitutes an upper bound for the dimer gap at wavevector $\Q$, 
the absence of a gap can simply be shown by demonstrating that either the numerator vanishes 
or the denominator diverges as a function of system size.
In the seminal paper by Rokhsar and Kivelson~\cite{Rokhsar88} they indeed used the fact that
the numerator vanishes at $(\pi,\pi)$ ($\hat{D}_x(\Q)$ with $\Q=(\pi,\pi)$ is a conserved quantity) to prove the gapless nature of the 
square lattice dimer spectrum at $(\pi,\pi)$. Similarly, one can prove the absence of a dimer gap at $(\pi,\pi,\pi)$
for the simple cubic lattice at the RK point~\cite{Huse03,Moessner03}. 
Furthermore, for square lattice again, Moessner and Sondhi~\cite{Moessner03} showed that the logarithmically divergent
structure factor $D_x(\Q)$ with $\Q=(\pi,0)$ leads via the SMA to another soft mode, which they called {\it pi0n}.

Another important property of the SMA is that it is equivalent to the {\em exact} first frequency
moment of the dynamical correlation function which is used to construct the trial excited state. 
In our case this is precisely the dimer density operator at wavevector $\Q$ and a 
given orientation. Then
\be
\omega^{D_x}_\mathrm{SMA}(\Q)
=\frac{1}{D_x(\Q)} \int_{0}^\infty\ \rmd \omega\ \omega\ \mathcal{D}_x(\Q,\omega) = \omega^{D_x}_\mathrm{fm}(\Q).
\ee
Historically the SMA bears its name from the approximation
\be
\mathcal{D}^\mathrm{SMA}_x(\Q,\omega)\approx D_x(\Q) \times \delta(\omega-\omega^{D_x}_\mathrm{SMA}(\Q)),
\ee
which is very sucessful for a large number of systems, most prominently for $^4$He as shown 
by Feynman~\cite{Feynman72}, but also for quantum Hall systems~\cite{Girvin85} and Haldane spin chains~\cite{Arovas88}. 
As an important result, we will show below that, for the QDMs, the SMA gives in general a rather poor account of the features 
contained in the dynamical dimer spectrum, with the exception of the hydrodynamic regime in the close neighborhood of the $(\pi,\pi)$ 
point on the square lattice and the $(\pi,\pi,\pi)$ point for the simple cubic lattice. 

\section{Numerical results for the dynamical dimer correlation functions}

In this section we present our numerical spectra for the various lattices. The Monte Carlo data has been obtained on systems
of linear size $L=64$ for the square lattice, $L=32$ for the cubic lattice and $L=36$ for the triangular lattice. 
For the spectral function measurements we used a uniform grid of $\Delta\tau=0.1$ up to $\tau_\mathrm{max}=80$ for the square and the 
cubic lattices. For the triangular lattice the required $\tau_\mathrm{max}$ value was smaller due to the finite gap.
For precise measurements of the $\tau$ derivative at the origin finer grids in $\Delta\tau$ have been used.

Our single plaquette flip dynamics conserves the winding number on the square lattice, the fluxes on the
cubic lattice and the topological sectors on the triangular lattice. We chose then to work in the zero winding number / zero flux 
sector for the square and cubic lattices respectively, while we worked in the trivial topological sector for the triangular 
lattice. For the square and the cubic lattice this choice might induce a slight bias towards the physics at $v/t<1$, since there the zero
winding number / flux sector is preferred. On the triangular lattice however the topological degeneracy insures that the results
in the different topological sectors are identical.


\subsection{Square Lattice}
\label{sec:square}
\begin{figure}
\begin{indented}
\item\includegraphics[width=\linewidth]{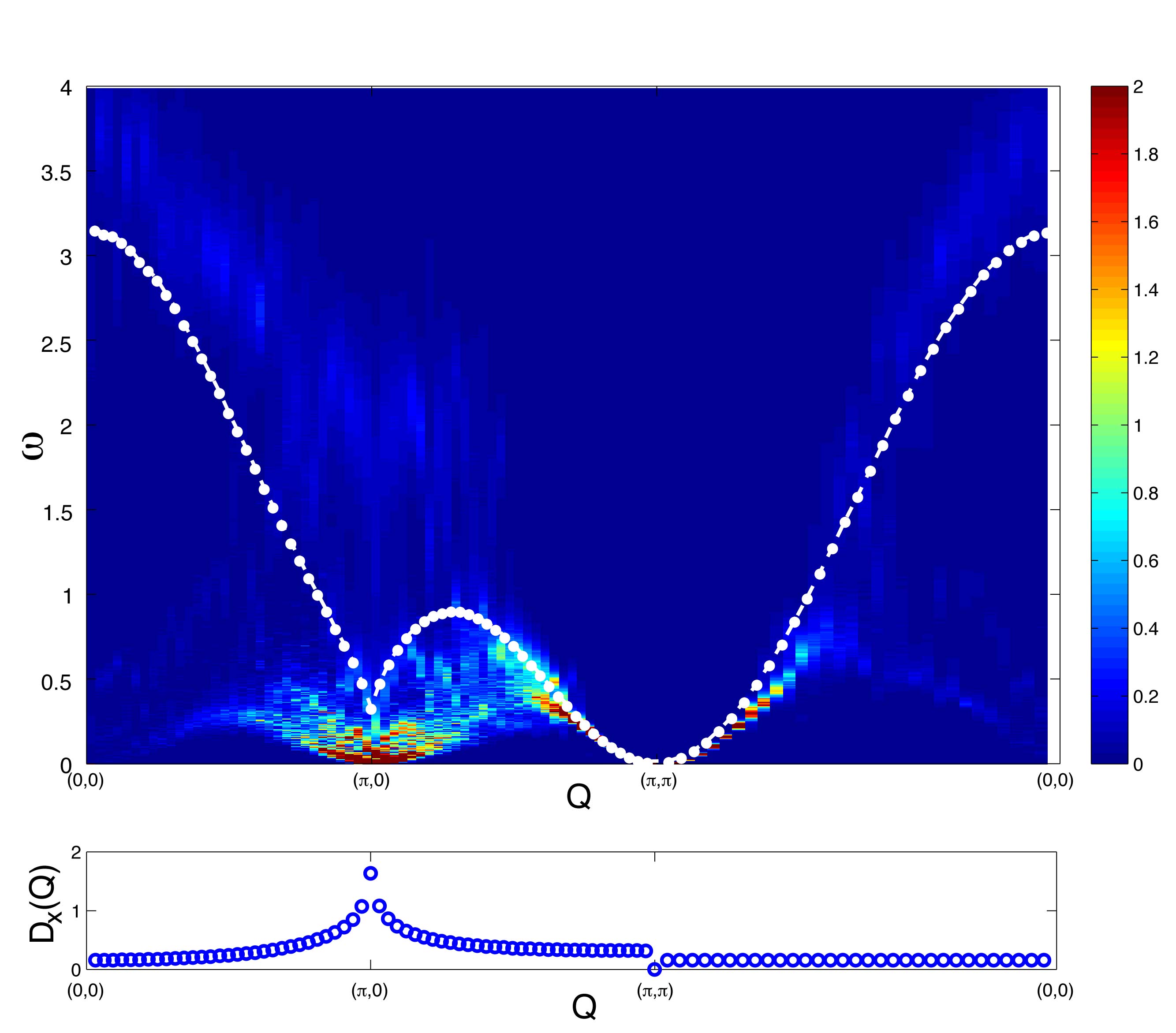}
\end{indented}
\caption{{\bf Square Lattice.}
Upper panel: dynamical dimer spectral functions $\mathcal{D}_x(\Q,\omega)$ 
plotted along the Brillouin zone path $(0,0)$-$(\pi,0)$-$(\pi,\pi)$-$(0,0)$ on a system of linear extension $L=64$.
The first moment $\omega^{D_x}_\mathrm{fm}(\Q)$ of the distribution  -- equivalent to the SMA prediction --
is shown by the white filled circles. Lower panel: equal-time dimer structure factor $D_x(\Q)$ along the same path.
The peak at $(\pi,0)$ is logarithmically divergent with system size.
\label{fig:overall_square}
}
\end{figure}
\begin{figure}
\begin{indented}
\item
\includegraphics*[width=\linewidth]{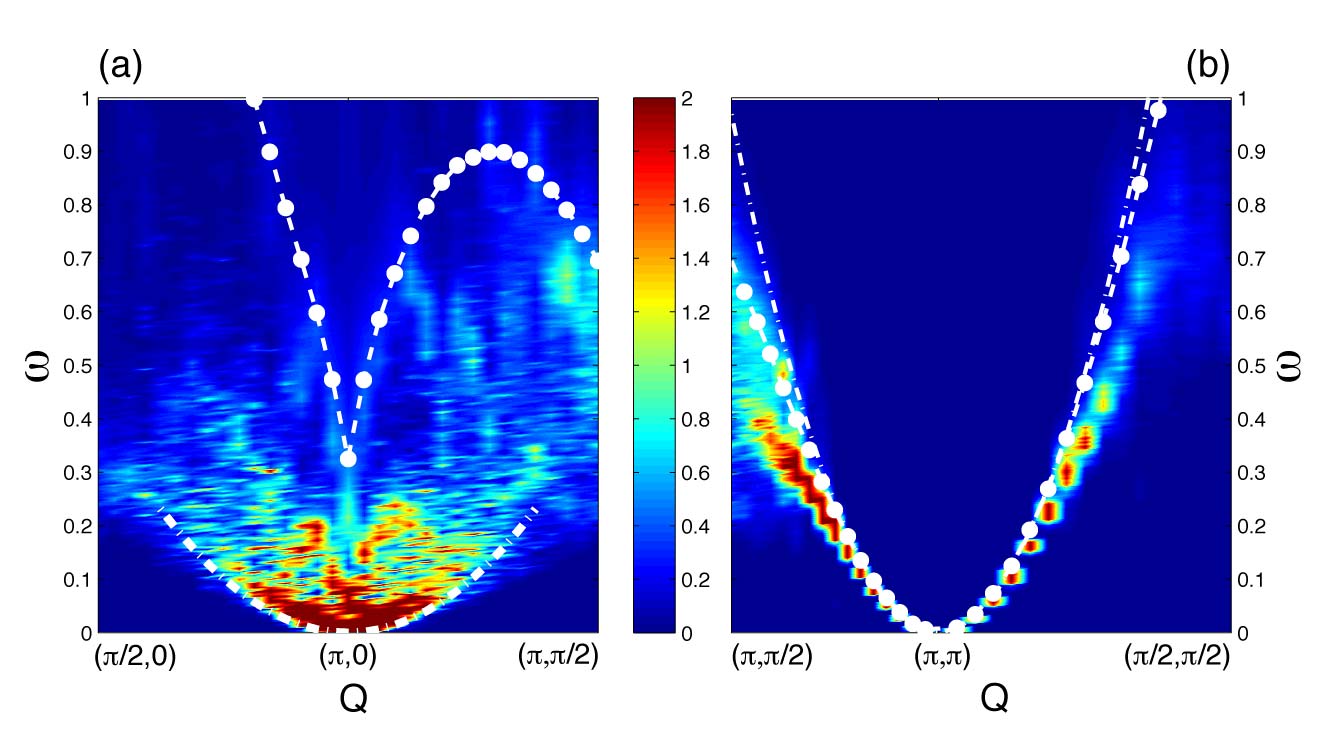}
\end{indented}
\caption{
{\bf Square Lattice.} 
Left panel: low-energy zoom of $\mathcal{D}_x(\Q,\omega)$ into the $(\pi,0)$ 
region. Right panel: similar zoom into the $(\pi,\pi)$ region. The dashed white
lines with the filled circles denotes the measured $\omega^{D_x}_\mathrm{fm}(\Q)$ for
both panels.  The dash-dotted line in the left panel is the analytical prediction 
\eref{eqn:pion_onset} for the maximum intensity at low energy, while in the right panel the dash-dotted
line shows the analytical prediction \eref{eqn:resonon_dispersion} for the resonon dispersion.
\label{fig:lowenergy_square}
}
\end{figure}

The RK point of the square lattice is a special point, since it is a critical point separating a 
staggered dimer phase at larger values of $v/t>1$ from a symmetry broken phase at 
$v/t<1$, whose detailed structure is still a topic of ongoing research~\cite{Fradkin04,Leung96,Syljuasen06,Ralko07b}. 

Since the seminal work of Rokhsar and Kivelson~\cite{Rokhsar88}, it is known that the excitation 
spectrum is gapless and has a quadratically dispersing dimer-wave mode becoming soft at $(\pi,\pi)$.
This SMA dispersion is given by~\cite{Rokhsar88,Perseguers06}:
\be
\omega_{\mathrm{resonon}}(\mathbf{k})=\frac{\pi}{8}k^2 + \ \mbox{higher order terms},
\label{eqn:resonon_dispersion}
\ee
where $\mathbf{Q}=\mathbf{k}+(\pi,\pi)$, and $k$ is the modulus of $\mathbf{k}$. It was realized
that $(\pi,\pi)$ is not the only gapless point in the Brillouin zone, when Ref.~\cite{Moessner03}
characterized a novel excitation termed {\em pi0n} which is gapless at $(\pi,0)$ and $(0,\pi)$, 
and which has its root in the height model description~\cite{Henley97}. 
However despite this knowledge of the "long-wavelength", low-energy behaviour, a full characterization of the dimer 
spectral response is lacking.

We present in the upper panel of Fig.~\ref{fig:overall_square} our numerical results for the dynamical 
dimer structure factor $\mathcal{D}_x(\Q,\omega)$ at the RK point on the square lattice obtained by the numerical
simulations outlined in \sref{sec:algo_obs}. The color highlights the intensity at a given $\Q$ and $\omega$, while the
line with the white circles denotes the first moment $\omega^{D_x}_{\mathrm{fm}}(\Q)$. In the lower panel the equal-time 
structure factor $D_x(\Q)$ is shown, which agrees nicely with the analytical results of \cite{Moessner03,Perseguers06}
(not shown). Note that since the parity of the number of dimers crossing any vertical line is conserved by the dynamics, 
$[D_x(\mathrm{q},\pi),H_{QDM}]=0$, which implies that the spectral weight vanishes along the
$\Q=(\mathrm{q},\pi)$ line, including the $(\pi,\pi)$ point. 
The non-continuity of the equal-time structure factor upon approaching $(\pi,\pi)$ from different directions
reflects the pinch (bow-tie) singularity in momentum space~\cite{Sandvik06}, and is a generic feature of two-dimensional 
bipartite Rokhsar-Kivelson points, giving rise to the critical real space dimer correlations. 
On the other hand the logarithmic divergence of $D_x(\pi,0)$ is specific to some 
models, since a lower stiffness $K$ in the low-energy action can lead to a non-divergent equal-time response, as 
is the case for example in a generalized classical model~\cite{Sandvik06}.
As expected the SMA prediction matches the full spectral function very well in the close vicinity of the $(\pi,\pi)$ point. However
in the rest of the Brillouin zone the intensity is by no means concentrated near the SMA result. We can however identify traces
of intensity above the SMA curve, loosely following it, for example in the neighborhood of $\Gamma=(0,0)$. 
On the other hand there is also some intensity at low frequencies and approaching zero energy as $\Q \rightarrow \Gamma$. 
Further investigation is needed to clarify whether this low energy signal is a "shadow band" feature of the $(\pi,0)$ or the
$(\pi,\pi)$ structure.

We now focus on the low-energy features around the $(\pi,0)$ and the $(\pi,\pi)$ region, which are presented in a zoomed 
view in the two panels of Fig.~\ref{fig:lowenergy_square}. First we briefly discuss the behavior in the $(\pi,\pi)$ region 
shown in the right panel. As stated before the SMA predicts a quadratically dispersing mode in the vicinity of this point.
Indeed our numerical results follow the refined analytical prediction~\eref{eqn:resonon_dispersion} presented by the dashed line 
very closely and the whole spectral function is actually dominated by the weight at this energy. 
It is difficult based on our numerical data for the spectral functions to address the linewidth of the mode, but based on the following
argument we expect a finite lifetime for this quadratic mode. As a general remark for a quadratic dispersion, it can easily be
understood that for any $\mathbf{Q}$, the gap should vanish in the thermodynamic limit. 
Indeed, by combining many low-lying excitations, such that the total momentum
is $\mathbf{Q}$, an arbitrary small energy state can be obtained. An equivalent statement is that a given $\mathbf{Q}$ 
excitation could decay into many lower-energy excitations (when there is no symmetry
preventing this process). Therefore, if we observe a quadratic behavior over a large region, this is because the spectral
weight on the lower energies states is tiny, due to density of states and matrix element effects.

The long-wavelength, long-time behavior of the $(\pi,0)$ feature has been 
discussed in Ref.~\cite{Moessner03}. Based on the analysis of the height model representation the asymptotic form
of the imaginary time correlation is known in the vicinity of $(\pi,0)$
\be
f(\mathbf{k},\tau) \sim \exp(-k\sqrt{\tau}), \label{eqn:longtau}
\ee
where now $\mathbf{Q}=\mathbf{k}+(\pi,0)$, and $k$ is the modulus of
$\mathbf{k}$.
This expression can be transformed by an inverse Laplace transform
giving:
\be
F(\mathbf{k},\omega)=\frac{k \exp(-\frac{k^2}{4\omega})}{2 \sqrt{\pi}\omega^{3/2}}.
\label{eqn:pion_freq}
\ee
The expression (\ref{eqn:longtau}) is supposed to be valid only at long times, 
as can be seen from the fact that the first moment of $F(\mathbf{k},\omega)$ 
does not exist due to the infinite $\tau$ derivative of $f(\mathbf{k},\tau)$ at $\tau=0$, while 
the true dynamical dimer structure factor has a bounded first moment for all $\mathbf{k}$.
Even though the first moment does not exist, it is possible to track the maximum
of $F(\mathbf{k},\omega)$ as a function of $\mathbf{k}$, giving the simple expression:
\be
\omega_{\mathrm{pi0n}}(\mathbf{k})=\frac{1}{6}k^2.
\label{eqn:pion_onset}
\ee
Note that the location of the maximum is also close to the onset of intensity in the 
formula \eref{eqn:pion_freq}. We plot this simple result by a dash-dotted line in the left panel 
of Fig.~\ref{fig:lowenergy_square}, and find very a nice agreement between this parameter-free prediction 
and the actual numerical results. The dimer excitation spectrum has definitely 
no quasiparticle peak in the vicinity of the $(\pi,0)$ point, but instead a characteristic broad continuum.

\subsection{Cubic Lattice}
\label{sec:cubic}
\begin{figure}
\begin{indented}
\item
\includegraphics[width=\linewidth]{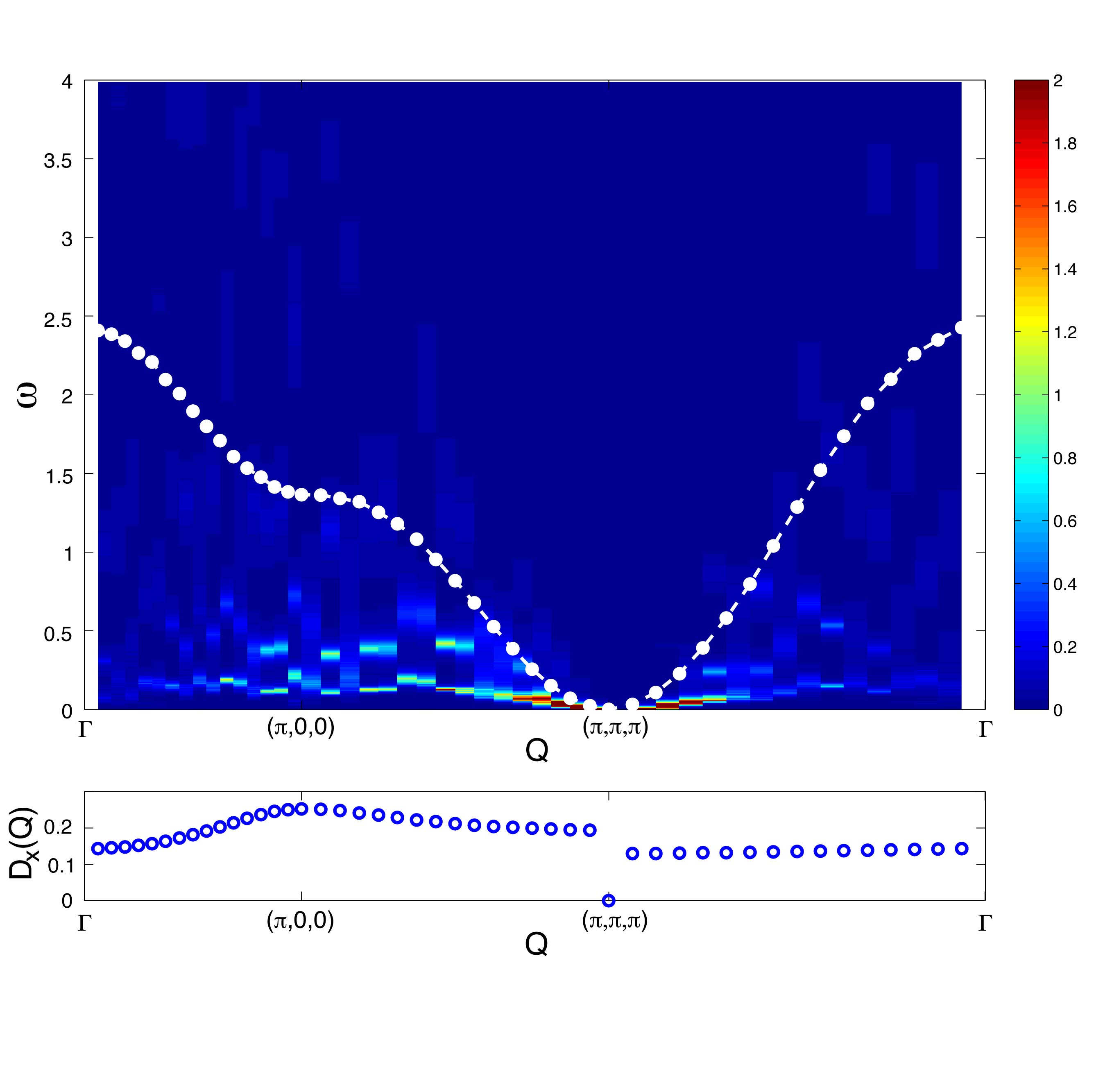}
\end{indented}
\caption{
{\bf Cubic Lattice.}
Upper panel: dynamical dimer spectral functions $\mathcal{D}_x(\Q,\omega)$ 
plotted along the Brillouin zone path $\Gamma=(0,0,0)\rightarrow(\pi,0,0)\rightarrow(\pi,\pi,\pi)\rightarrow\Gamma$ 
on a system of linear extension $L=32$.
The first moment $\omega^{D_x}_\mathrm{fm}(\Q)$ of the distribution  -- equivalent to the SMA prediction --
is shown by the white filled circles. Lower panel: equal-time dimer structure factor $D_x(\Q)$ along the same path.
Note that $D_x(\Q)$ has no divergent feature on the cubic lattice.
\label{fig:overall_cubic}
}
\end{figure}
\begin{figure}
\begin{indented}
\item
\includegraphics[width=\linewidth]{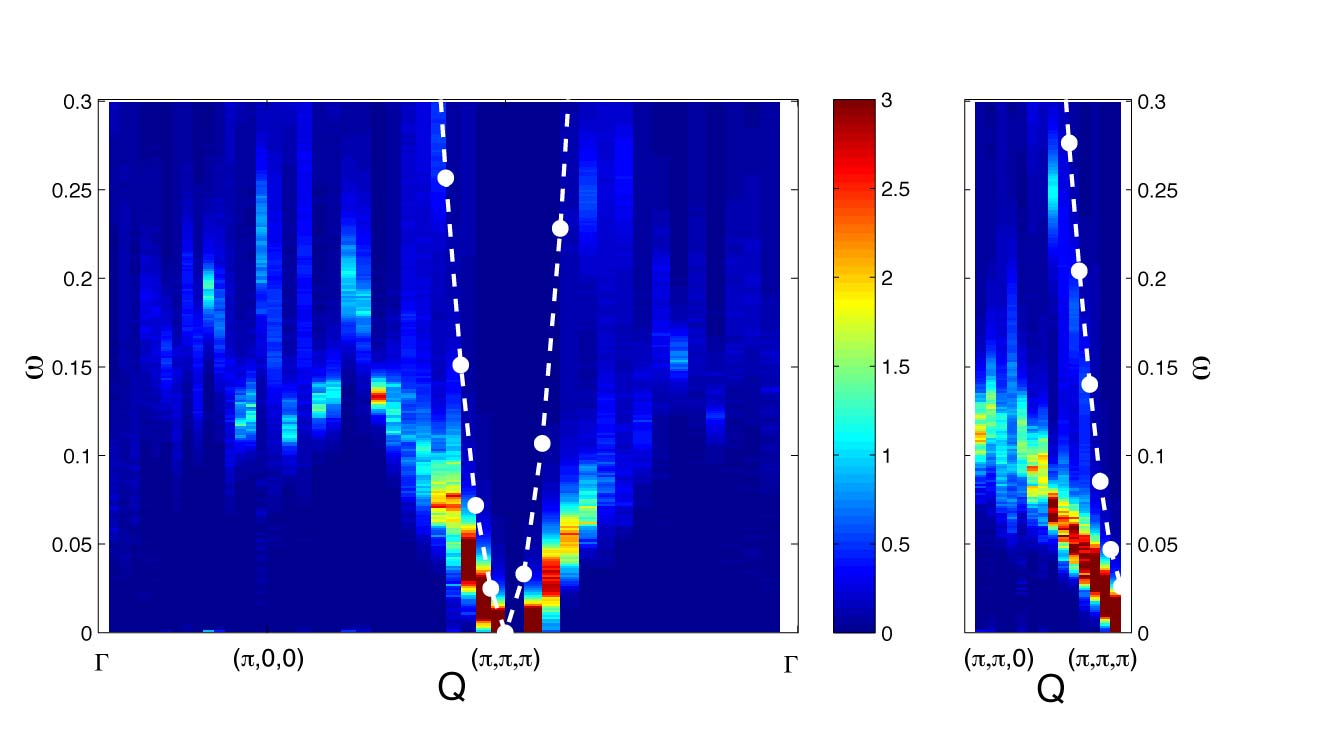}
\end{indented}
\caption{
{\bf Cubic Lattice.}
Left panel: low-energy zoom $\omega\in [0,0.3]$ of $\mathcal{D}_x(\Q,\omega)$ 
along the same path in the Brillouin zone as in \Fref{fig:overall_cubic}. 
Right panel: Supplementary low energy data for $\mathcal{D}_x(\Q,\omega)$ along the path
$(\pi,\pi,0) \rightarrow (\pi,\pi,\pi)$.
The dashed white lines with the filled circles denotes the measured $\omega^{D_x}_\mathrm{fm}(\Q)$ for
both panels, and is in good agreement with the expected $k^2$ dispersion close to $(\pi,\pi,\pi)$.
Note that the maximum intensity of $\mathcal{D}_x(\Q,\omega)$ rapidly deviates from the expected $k^2$-law 
and becomes linear, revealing the remnant of the linearly dispersing "photon" mode of the $U(1)$-liquid 
Coulomb phase.
\label{fig:lowenergy_cubic}
}
\end{figure}

We now turn to the three-dimensional (3d) cubic lattice. Being also
bipartite, a well defined mapping to a long wavelength action exists~\cite{Huse03}, 
and this model is expected to be somewhat similar to the two-dimensional 
square lattice case. In particular the SMA argument predicts a quadratic dispersion
close to the $(\pi,\pi,\pi)$ point~\cite{Moessner03}. 
However, contrary to the square lattice case, there is no other peculiar point in the 
Brillouin zone (such as the 2d "pi0n") so that the $(\pi,\pi,\pi)$ point is supposed to be 
the only gapless excitation.
Also, in the 3d case, there is a so-called Coulomb phase
with dipolar dimer correlations~\cite{Huse03,Moessner03,Fradkin04}.
Moreover, based on an effective low energy field-theory~\cite{Huse03,Moessner03}, the
cubic dimer model can be described: (a) for $v_c/t<v/t<1$, by a standard U(1) 
Maxwell electromagnetic theory with a linearly dispersing transverse \emph{photon}; 
(b) at the RK point, $v=t$, the photon dispersion becomes quadratic.  

Fig.~\ref{fig:overall_cubic} shows our results both for the dimer
spectral functions and equal-time structure factors along a specific path through
the Brillouin zone. Except at $(\pi,\pi,\pi)$ where it vanishes by
symmetry, the equal-time structure factor $D_x(\Q)$ is rather smooth and
has no divergence. The non-continuity of $D_x(\Q)$ upon approaching
$(\pi,\pi,\pi)$ from different directions is the momentum space signature of the 
dipolar real-space dimer correlation functions.

The first moment of the spectral functions, which corresponds to the SMA result, exhibits a wide 
dispersion reaching its maximum of $\sim$ 2.45 at the $\Gamma$ point, with a quadratic behavior 
aproaching zero energy close to $(\pi,\pi,\pi)$, similar to the square lattice.
The visible intensity is however concentrated mostly at very low energy, in contrast to the square lattice.
By looking more closely at the low-energy features shown in Fig.~\ref{fig:lowenergy_cubic}, it can be seen 
that  the spectral function is not simply following the SMA energy, even in the neighorhood of $(\pi,\pi,\pi)$, 
clearly at variance with the  results for the square lattice, where the intensity was well described by the SMA 
prediction near $(\pi,\pi)$, see the right panel of Fig.~\ref{fig:lowenergy_square}. 
Instead we see a {\em linearly} dispersing feature with an approximate velocity $c\approx 0.04(1)$ in all directions 
explored around $(\pi,\pi,\pi)$. Of course, such a linear behavior cannot exist arbitrary close to $(\pi,\pi,\pi)$, 
because the SMA correctly predicts a quadratic dispersion, but nothing prevents such a linear mode with a tiny velocity
further away from this point, i.e. for energies above a certain crossover scale. This unexpected feature probably has its origin in the 
truly linearly dispersing photon mode expected at smaller $v/t$ adjacent to the RK point, and whose
velocity is supposed to grow as a function of $(1-v/t)$.
Our observation calls for an investigation of the cubic QDM using Green's function Monte Carlo techniques in order to
explore the dispersion of the photon deep inside the Coulomb phase and when approaching the RK point from smaller $v/t<1$ values.


\subsection{Triangular Lattice}
\label{sec:triangular}
\begin{figure}
\begin{indented}
\item
\includegraphics[width=\linewidth]{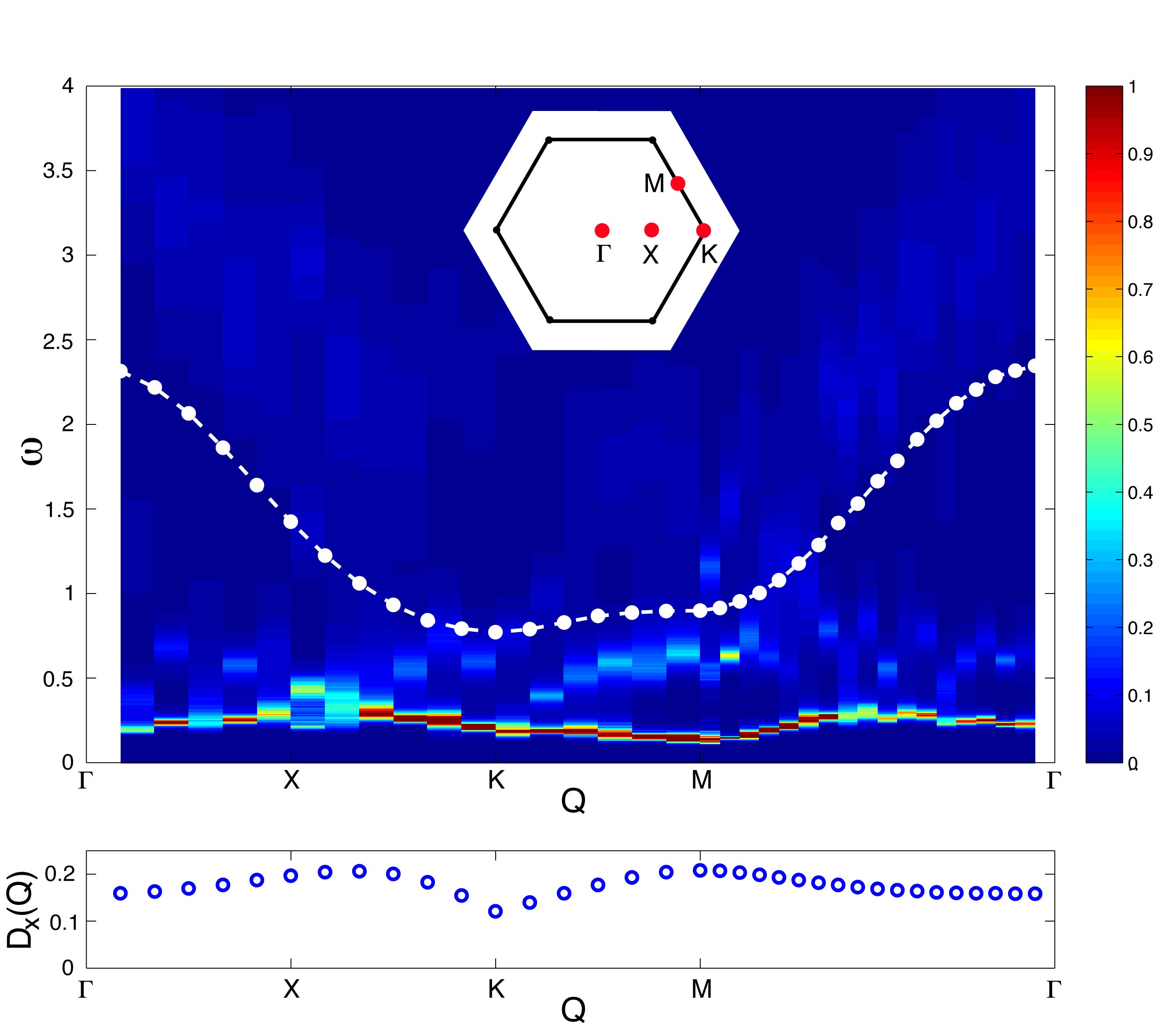}
\end{indented}
\caption{
{\bf Triangular Lattice.}
Upper panel: dynamical dimer spectral functions $\mathcal{D}_x(\Q,\omega)$ 
plotted along the Brillouin zone path: $\Gamma \rightarrow X \rightarrow K \rightarrow M \rightarrow \Gamma$ 
for a system of linear extension $L=36$. The points in the Brillouin zone are labeled in the inset. 
The first moment $\omega^{D_x}_\mathrm{fm}(\Q)$ of the distribution  
-- equivalent to the SMA prediction -- is shown by the white filled circles. 
Lower panel: equal-time dimer structure factor $D_x(\Q)$ along the same path. 
\label{fig:overall_triangular}
}
\end{figure}
\begin{figure}
\begin{indented}
\item
\includegraphics[width=\linewidth]{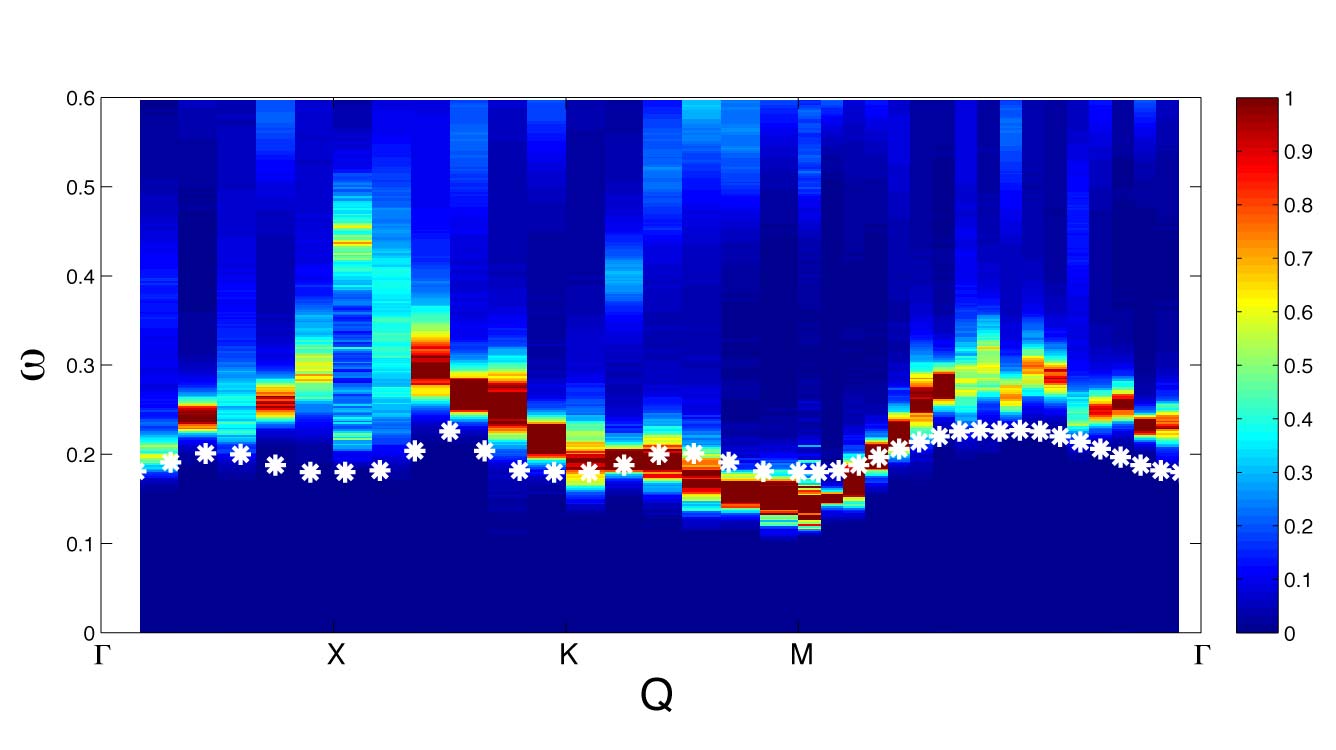}
\end{indented}
\caption{
{\bf Triangular Lattice.}
 Low-energy part of the dynamical dimer spectral function
 along the same path in the Brillouin zone as in \Fref{fig:overall_triangular}.  
The onset of the two-vison continuum is indicated by the white stars.
The single vison energies used here have been obtained in Ref.~\cite{Ivanov04}.
\label{fig:lowenergy_triangular}
}
\end{figure}
\begin{figure}
\begin{indented}
\item
\includegraphics[width=0.9\linewidth]{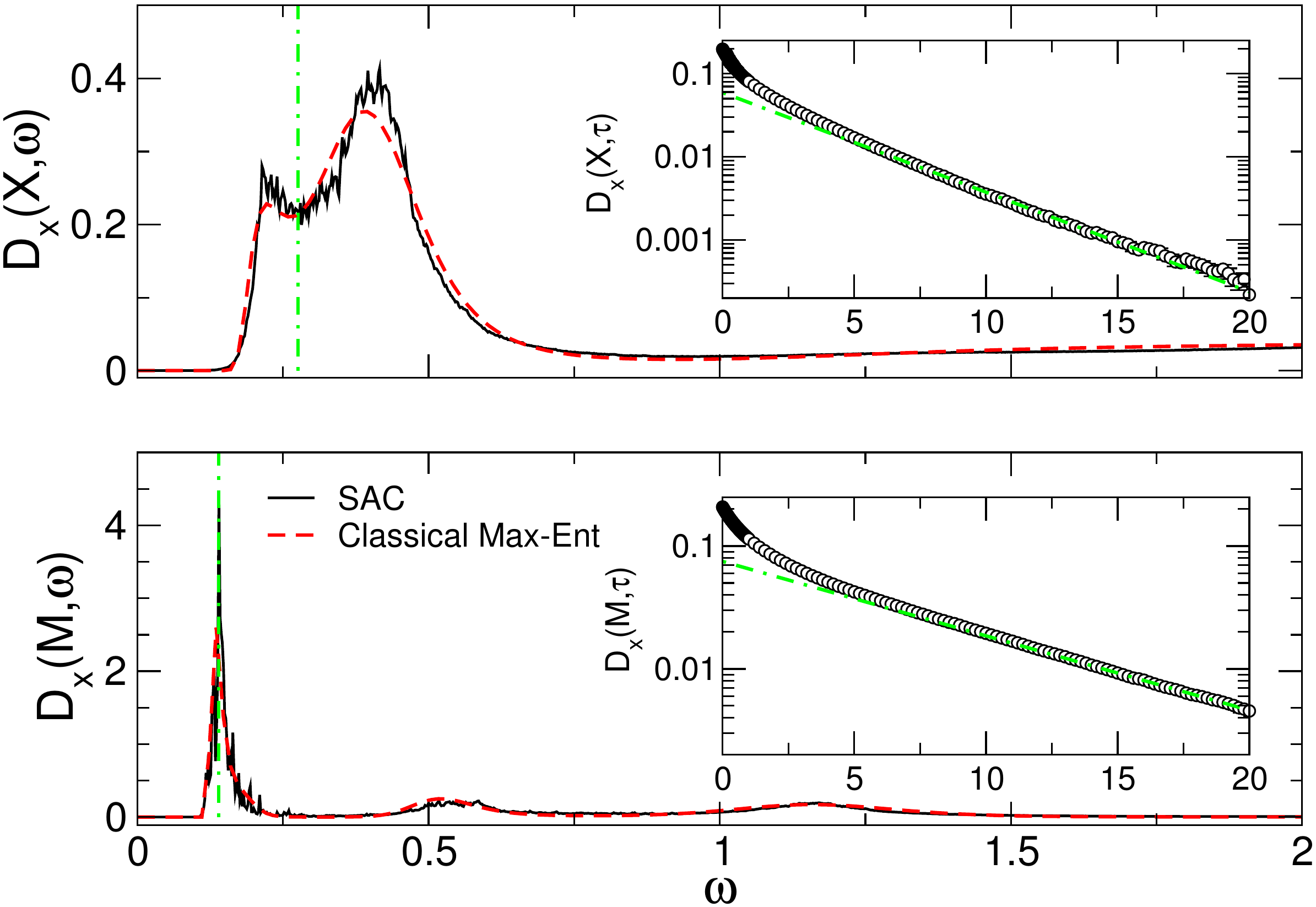}
\end{indented}
\caption{
{\bf Triangular Lattice.}
Comparison of Classical Maximum-Entropy and Stochastic analytical continuation spectra for the dimer spectral functions
at the $X$ and the $M$ points. Both methods for analytic continuation yield very similar spectra.
The vertical green dash-dotted lines indicate the dimer gap energies extracted in Ref.~\cite{Ralko06}.
Insets: Monte Carlo imaginary time data used for the analytical continuation. The dash-dotted green 
line indicates the fit with a single exponential decay obtained in~\cite{Ralko06}.
It is obvious that the spectral function at the $X$-point has no sizable quasiparticle peak, while the
spectral function at the $M$-point seems to show a resolution limited peak at the lowest energy, providing evidence
for a coherent dimer excitation at this momentum.
 \label{fig:spectralfunctions_triangular}
}
\end{figure}

In contrast to the previous two bipartite lattices, the triangular lattice QDM at the RK point is 
known to be in a gapped $Z_2$-liquid phase sustaining topological order~\cite{Moessner01}, 
which is is stable in a finite region of parameter space for $0.8 \lesssim v/t \le 1$. The 
dimer excitations possess a finite gap in the whole Brillouin zone~\cite{Ivanov04}. 

Our numerical data shown in Fig.~\ref{fig:overall_triangular} clearly reveal the presence of a 
finite gap. Even more striking is the fact that the first spectral moment $\omega^{D_x}_\mathrm{fm}(\Q)$ 
is nowhere in the Brillouin zone close to the the intensity carrying low energy excitations, and even the
location of the minimum of the SMA predication ($K$ point) and the actual lowest dimer excitation 
($M$ point) do not coincide. This is in contrast to the square and cubic lattice, where the SMA gave a faithful 
description of the dimer exciation energies at the $(\pi,\pi)$ and the $(\pi,\pi,\pi)$ point respectively.
The equal-time structure factor is smooth on the triangular lattice, due to the short-ranged nature of the dimer correlations in real space.

Looking more closely at the low energy part of the spectral functions in Fig.~\ref{fig:lowenergy_triangular}
one recognizes that the intensity carrying modes at the onset of the dimer spectrum are much less
dispersive than the SMA prediction suggests. Furthermore the broadening can vary through the Brillouin zone.
For instance, the spectrum is very broad at the $X$ point, while it is much sharper close to the $M$ point, 
see Fig.~\ref{fig:spectralfunctions_triangular}.
These observations strongly suggest that the dimer waves excited by the operator \eref{eqn:resonon_structure} are 
not the true "elementary" excitations of this dimer model. 

Indeed the elementary excitations are known to consist of so-called \emph{visons} which are non-local in
terms of dimers and for which the dispersion has been computed~\cite{Ivanov04}. Real dimer excitations can 
be understood as 2- or more vison excitations. Here, the lowest-energy dimer excitation (at the $M$ point)
is formed by a bound-state of two visons~\cite{Ivanov04}. Based on the single vison dispersion relation~\cite{Ivanov04}, 
we have plotted in Fig.~\ref{fig:lowenergy_triangular} the bottom of the 2-vison continuum. 
By comparing to the spectral weight of the dimer correlations, we observe that: (a) close to $K$
and $\Gamma$ points, the spectral weight nicely follows the 2-vison continuum; (b) the lowest energy 
state at $M$ is slightly below the two-vison gap, which is consistent with the interpretation as a two-vison bound-state; 
(c) at the $X$ point, the spectral weight is very broad, we cannot
reveal any well-formed excitation so that a dimer excitation probably disintegrates in the 2-vison
continuum, a point of view which has also been raised in Refs.~\cite{Ivanov04,Ralko07}. These observations then suggest
a scenario where the coherent (quasiparticle-like) dimer excitations close to the $M$ point cross into the continuum going either to
the $K$ or the $\Gamma$ point, giving rise to damped, incoherent excitations for example at the $X$ point.
It is also possible that there are termination points similar to magnetic systems or liquid Helium 4~\cite{Stone06}.

A natural extension of this work would be to investigate the spectral functions of the visons directly, first at the RK point~\cite{Ivanov04}, 
and then also towards the phase transition from the topologically ordered phase into the symmetry broken $\sqrt{12}\times \sqrt{12}$ 
phase~\cite{Ralko07}.
It will be interesting to confirm that the visons have a sizable quasiparticle weight in the topologically ordered phase, and therefore can really
be considered as the elementary excitations inside this phase. 
Also the currently unexplored nature of the excitations above the $\sqrt{12}\times \sqrt{12}$ ordered groundstate 
(dimers vs visons) could be 
clarified this way.

\section{Conclusions and perspectives}
\label{sec:conclusion}

Quantum dimer models at the Rokhsar-Kivelson point can be efficiently simulated with a classical 
Monte-Carlo algorithm and in particular, imaginary time-displaced dimer correlations can be obtained. 
By performing a stochastic analytic continuation to real frequencies, we have been able to compute
quantum dimer spectral functions for a variety of lattices. 

For bipartite lattices (2d square and 3d cubic), we have confirmed that these models are in a gapless phase, with a rather well-defined resonon excitation dispersing quadratically. The 2d case also exhibits 
other gapless excitations close to $(\pi,0)$ (pi0ns), that are not described by a simple dispersion 
but rather a broad spectrum in full agreement with analytic calculations. In the 3d case, on top of the
quadratic dispersing mode, we have shown evidence that a linear-dispersing excitation might play 
a role at shorter wavelengths and might be related to the well-known photon-like excitation that appears
away from the RK point. 

For non-bipartite lattices (2d triangular), these models are in a $Z_2$-liquid phase with a finite-gap 
for dimer excitations. By analyzing the low-energy structures, we have provided arguments that, 
for different points in the Brillouin zone, the lowest energy dimer excitations can be 
viewed either as a coherent two-vison bound-state, or on the contrary can decay into the two-vison continuum.

Similar future studies could investigate further open questions, just to cite a few~: 
\begin{enumerate}
\item 
analytic continuation of Green's function Monte Carlo data away from the 
RK point (the possibility to approximately compute time-displaced correlations 
has already been used in Ref.~\cite{Ralko06}), 
observation of the linearly dispersing "photon" of $U(1)$ gauge theories on 
3d bipartite lattices~\cite{Huse03,Moessner03,Fradkin04}.
\item
spectral functions and coherence of the visons on the triangular lattice. 
\item 
The stiffness appearing in the low-energy approach can be varied by changing the model. 
In Ref.~\cite{Alet06}, it has been shown that, by adding an aligning term, the stiffness can be increased continuously from 
1/2 to 4, before a phase transition occurs. In Ref.~\cite{Sandvik06}, by including longer-range bipartite dimers, the stiffness
changes smoothly from 1/2 to 1/9, while allowing for  non-bipartite dimers immediately opens a gap. 

\end{enumerate}

\ack
We thank F.~Alet, C.L.~Henley, D.A.~Ivanov, F.~Mila, G.~Misguich, R.~Moessner, and A.~Ralko 
for valuable discussions. We also thank D.A.~Ivanov for providing us with his vison gap data.
The simulations have been performed with a C++ code based on some of the ALPS 
libraries \cite{Alps05}. We benefited from allocations of CPU time on the Cray XT3 at the CSCS in Manno.


\section*{References}
\begin{thebibliography}{99}
\bibitem{Moessner01} R.~Moessner and S.L.~Sondhi, 
{\it Resonating Valence Bond Phase in the Triangular Lattice Quantum Dimer Model},
Phys. Rev. Lett {\bf 86}, 1881 (2001).
\bibitem{Moessner01a} R. Moessner, S. L. Sondhi, and E. Fradkin, 
{\it Short-ranged resonating valence bond physics, quantum dimer models, and Ising gauge theories}, 
Phys. Rev. B {\bf 65}, 024504 (2001).
\bibitem{Misguich02}
G.~Misguich, D.~Serban, and V.~Pasquier,
{\it Quantum Dimer Model on the Kagome Lattice: Solvable Dimer-Liquid and Ising Gauge Theory},
Phys. Rev. Lett. 89, 137202 (2002).
\bibitem{Ivanov04}
D.A.~Ivanov,
{\it Vortexlike elementary excitations in the Rokhsar-Kivelson dimer model on the triangular lattice},
Phys. Rev. B {\bf 70}, 094430 (2004).
\bibitem{Huse03}
D.A.~Huse, W.~Krauth, R.~Moessner and S. L.~Sondhi,
{\it Coulomb and liquid dimer models in three dimensions},
Phys. Rev. Lett. {\bf 91}, 167004 (2003).
\bibitem{Moessner03}
R.~Moessner and S.L.~Sondhi,
{\it Three-dimensional resonating-valence-bond liquids and their excitations},
Phys. Rev. B {\bf 68}, 184512 (2003).
\bibitem{Rokhsar88}
D.S.~Rokhsar and S.A.~Kivelson,
{\it Superconductivity and the Quantum Hard-Core Dimer Gas},
Phys. Rev. Lett. {\bf 61}, 2376 (1988).
\bibitem{Balents05}
L. Balents, L. Bartosch, A. Burkov, S. Sachdev, and K. Sengupta, 
{\it Putting competing orders in their place near the Mott transition. II. The doped quantum dimer model}, 
 Phys. Rev. B {\bf 71}, 144509 (2005).
\bibitem{Poilblanc06} 
D. Poilblanc, F. Alet, F. Becca, A. Ralko, F. Trousselet, and F. Mila, 
{\it Doping quantum dimer models on the square lattice}, 
 Phys. Rev. B {\bf 74}, 014437 (2006).
\bibitem{Ralko06} 
A. Ralko, M. Ferrero, F. Becca, D.  Ivanov, and F. Mila, 
{\it Dynamics of the quantum dimer model on the triangular lattice: Soft modes and local resonating valence-bond correlations},
Phys. Rev. B {\bf 74}, 134301 (2006).
\bibitem{Ralko07}
A. Ralko, M. Ferrero, F. Becca, D.  Ivanov, and F. Mila, 
{\it Crystallization of the resonating valence bond liquid as vortex condensation},
Phys. Rev. B {\bf 76}, 140404(R) (2007).
\bibitem{Blote82} W. J. Bl\"ote and H. J. Hilhorst, 
{\it Roughening transitions and the zero-temperature triangular Ising antiferromagnet}, 
J. Phys. A {\bf 15}, L631 (1982).
\bibitem{Henley97}
C.L. Henley,
{\it Relaxation time for a dimer covering with height represenation},
J. Stat. Phys. {\bf 89}, 483 (1997). 
\bibitem{Fendley02} P. Fendley, R. Moessner, and S. L. Sondhi,
{\it Classical dimers on the triangular lattice},
 Phys. Rev. B {\bf 66}, 214513 (2002).
\bibitem{Ioselevich2002} A. Ioselevich, D.A. Ivanov, and M.V. Feigelman, 
{\it Ground-state properties of the Rokhsar-Kivelson dimer model on the triangular lattice},
Phys. Rev. B {\bf 66}, 174405 (2002).
\bibitem{Henley04}
C.L.~Henley,
{\it From classical to quantum dynamics at Rokhsar-Kivelson points},
J. Phys.: Condens. Matter {\bf 16}, S891 (2004).
\bibitem{Syljuasen05}
O.F.~Syljuasen,
{\it Continuous-time diffusion Monte Carlo method applied to the quantum dimer model},
Phys. Rev. B {\bf 71}, 020401 (2005).
\bibitem{Moessner03b}
R. Moessner and S.L. Sondhi,
{\it Theory of the {[}111{]} magnetization plateau in spin ice},
Phys. Rev. B {\bf 68}, 064411 (2003).
\bibitem{Cabra04}
D.C. Cabra, M.D. Grynberg, P.C.W. Holdsworth, A. Honecker, P. Pujol, J. Richter, D. Schmalfuss, J. Schulenburg,
{\it Quantum kagome antiferromagnet in a magnetic field: Low-lying non-magnetic excitations versus valence-bond crystal order},
Phys. Rev. B {\bf 71}, 144420 (2005).
\bibitem{Bergman06}
D.L. Bergman, R. Shindou, G.A. Fiete, and L. Balents,
{\it Quantum Effects in a Half-Polarized Pyrochlore Antiferromagnet},
Phys. Rev. Lett. {\bf 96}, 097207 (2006).
\bibitem{MaxEnt}
M. Jarrell and J. E. Gubernatis, 
{\it Bayesian inference and the analytic continuation of imaginary-time quantum Monte Carlo data},
Physics Report {\bf 269}, 113 (1996).
\bibitem{Sandvik98}
A.W.~Sandvik, 
{\it Stochastic method for analytic continuation of quantum Monte Carlo data},
Phys. Rev. B {\bf 57},  10287  (1998).
\bibitem{Beach04a}
K.S.D.~Beach, 
{\it Identifying the maximum entropy method as a special limit of stochastic analytic continuation},
preprint \verb+cond-mat/0403055+ (2004).
\bibitem{Feynman72}
R.P. Feynman,
{\it Statistical Mechanics},
(Benjamin, Reading, Mass. 1972), Chap. 11, and references therein.
\bibitem{Girvin85}
S.M. Girvin, A.H. MacDonald, and P.M. Platzman,
{\it Collective-Excitation Gap in the Fractional Quantum Hall Effect},
Phys. Rev. Lett. {\bf 54}, 581 (1985).
\bibitem{Arovas88}
D.P. Arovas, A. Auerbach, and F.D.M. Haldane,
{\it Extended Heisenberg models of antiferromagnetism: Analogies to the fractional quantum Hall effect},
Phys. Rev. Lett. {\bf 60}, 531 (1988).
\bibitem{Fradkin04}
E. Fradkin, D.A. Huse, R. Moessner, V. Oganesyan, and S.L. Sondhi, 
{\it Bipartite Rokhsar-Kivelson points and Cantor deconfinement},
Phys. Rev. B {\bf 69}, 224415 (2004).
\bibitem{Leung96}
P.W. Leung, K.C. Chiu, and K.J. Runge,
{\em Columnar dimer and plaquette resonating-valence-bond orders in the quantum dimer model},
Phys. Rev. B {\bf 54} 12938 (1996).
\bibitem{Syljuasen06}
O.F. Syljuasen,
{\em Plaquette phase of the square-lattice quantum dimer model: Quantum Monte Carlo calculations},
Phys .Rev. B {\bf 73} 245105 (2006).
\bibitem{Ralko07b}
A. Ralko, D. Poilblanc, and R. Moessner,
{\it Generic mixed columnar-plaquette phases in Rokhsar-Kivelson models},
preprint \verb+arXiv:0710.1269+ (2007).
\bibitem{Perseguers06} 
S. Perseguers,
{\it Quantum Dimer Model: from Triangular to Square Lattice and Vice Versa},
Diploma thesis,
EPF Lausanne, Switzerland, (2006).
\bibitem{Sandvik06}
A.W.~Sandvik and R.~Moessner,
{\it Correlations and confinement in nonplanar two-dimensional dimer models},
Phys. Rev. B {\bf 73}, 144504 (2006).
\bibitem{Stone06}
M.B.~Stone, I.A.~Zaliznyak, T.~Hong, C.L.~Broholm, and D.H.~Reich,
{\it Quasiparticle breakdown in a quantum spin liquid},
Nature {\bf 440}, 187 (2006).
\bibitem{Alet06} F. Alet, J. L. Jacobsen, G. Misguich, V. Pasquier, F. Mila and M. Troyer, 
{\it Interacting classical dimers on the square lattice},
Phys. Rev. Lett. {\bf 94}, 235702 (2005).
\bibitem{Alps05}
F. Albuquerque et al. (ALPS collaboration),
{\it The ALPS project release 1.3: open source software for strongly correlated systems},
J. Magn. Mag. Mat. {\bf 310}, 1187 (2007);
M. Troyer, B. Ammon and E. Heeb, 
{\it Parallel Object Oriented Monte Carlo Simulations},
Lecture Notes in Computer Science, {\bf 1505}, 191 (1998). 
\end{thebibliography}
\end{document}